\documentclass[aps,prl,twocolumn,superscriptaddress,showpacs,longbibliography]{revtex4-1}
\usepackage[colorlinks=true,citecolor=blue,linkcolor=blue,breaklinks=true]{hyperref}
\usepackage{amssymb}
\usepackage{graphicx}
\usepackage{amsmath}

\usepackage{times}
\usepackage{color}
\usepackage{subfigure}
\usepackage{setspace}
\usepackage{bm}

\begin{document}

\newcommand {\beq} {\begin{equation}}
\newcommand {\eeq} {\end{equation}}
\newcommand {\bqa} {\begin{eqnarray}}
\newcommand {\eqa} {\end{eqnarray}}
\newcommand {\ba} {\ensuremath{b^\dagger}}
\newcommand {\Ma} {\ensuremath{M^\dagger}}
\newcommand {\psia} {\ensuremath{\psi^\dagger}}
\newcommand {\psita} {\ensuremath{\tilde{\psi}^\dagger}}
\newcommand{\lp} {\ensuremath{{\lambda '}}}
\newcommand{\A} {\ensuremath{{\bf A}}}
\newcommand{\Q} {\ensuremath{{\bf Q}}}
\newcommand{\kk} {\ensuremath{{\bf k}}}
\newcommand{\qq} {\ensuremath{{\bf q}}}
\newcommand{\kp} {\ensuremath{{\bf k'}}}
\newcommand{\rr} {\ensuremath{{\bf r}}}
\newcommand{\rp} {\ensuremath{{\bf r'}}}
\newcommand {\ep} {\ensuremath{\epsilon}}
\newcommand{\nbr} {\ensuremath{\langle ij \rangle}}
\newcommand {\no} {\nonumber}
\newcommand{\up} {\ensuremath{\uparrow}}
\newcommand{\dn} {\ensuremath{\downarrow}}
\newcommand{\rcol} {\textcolor{red}}

\begin{abstract}
We study the dynamics of one and two dimensional disordered lattice bosons/fermions
initialized to a Fock state with a pattern of $1$ and $0$
particles on $A$ and ${\bar A}$ sites. For non-interacting systems we
establish a universal relation between the long time density
imbalance between $A$ and ${\bar A}$ site, $I(\infty)$, the localization length $\xi_l$, and the geometry of the initial
pattern. For alternating initial pattern of $1$ and $0$ particles in 1
dimension, $I(\infty)=\tanh[a/\xi_l]$, where $a$ is the lattice
spacing. For systems with mobility edge, we find
analytic relations between $I(\infty)$, the effective localization length $\tilde{\xi}_l$ and the fraction of localized states
$f_l$. The imbalance as a function of disorder shows non-analytic behaviour when the mobility edge passes through a band edge. For
interacting bosonic systems, we show that dissipative processes lead to a
decay of the memory of initial conditions. However, the excitations
created in the process act as a bath, whose noise correlators retain
information of the initial pattern. This sustains a finite imbalance
at long times in strongly disordered interacting systems.

 \end{abstract}
\title{Memories of initial states and density imbalance in dynamics of
  interacting disordered systems}
\author{Ahana Chakraborty}\email{ahana@theory.tifr.res.in}
 \affiliation{Department of Theoretical Physics, Tata Institute of Fundamental
 Research, Mumbai 400005, India.}
\author{ Pranay Gorantla} 
\affiliation{Department of Theoretical Physics, Tata Institute of Fundamental
 Research, Mumbai 400005, India.}
\affiliation{Department of Physics, Princeton University, Washington Road,
Princeton, NJ 08544, USA}
\author{Rajdeep Sensarma}
 \affiliation{Department of Theoretical Physics, Tata Institute of Fundamental
 Research, Mumbai 400005, India.}

\date{\today}

\maketitle

A generic quantum many body system, initialized to a typical state,
forgets the memory of the initial state. In the long time limit,
local observables in the system can be described by an ensemble of states with a
probability measure determined by its Hamiltonian. This basic tenet of
equilibrium statistical mechanics has been challenged in recent years
in strongly disordered interacting
quantum systems, a phenomenon called many body
localization(MBL)~\cite{BaskoAleiner,Mirlin2005,pal_huse, MBL_review1,MBL_review2,MBL_review3}.

While theoretical studies of MBL have focussed on the
properties of many body eigenstates in the middle of the
spectrum~\cite{MBL_review1,MBL_review2,MBL_review3,pal_huse}, it is impossible to
experimentally access these states individually. 
In experiments on MBL in cold atoms~\cite{MBLexpt1,MBLexpt2,MBLexpt3,MBLexpt4,MBLexpt5,MBLexpt11}, the system is initialized in a Fock
state, which has $1$ particle on a set of lattice sites (say $A$) and $0$
particles on the rest (say $\bar{A}$). As the system
evolves, the density imbalance between $A$ and $\bar{A}$
sites, normalized by average density, is measured. The Hamiltonian of the system (averaged
over disorder) does not distinguish between $A$ and $\bar{A}$
sites; in a thermal state the imbalance should be $0$. A finite imbalance in
the long time limit implies that the system remembers the initial
condition and indicates absence of thermalization in the
system.

In this paper, we use a new extension of Keldysh field theory
~\cite{chakraborty2} to
understand imbalance dynamics in systems with random or
incommensurate potentials. For localized non-interacting systems, we derive a universal relation between the long time
density imbalance, the localization length and the geometry of the initial
density pattern. For the initial patterns used in 1-d and 2-d
experiments, we obtain analytic relations between localization
length and long time imbalance. Near a localization-delocalization
transition, the imbalance scales as the inverse localization
length. We test our theory using the
random potential Anderson model~\cite{AL_original} in 1 and 2 dimensions and the Aubry Andre model~\cite{AA_original} in 1
dimension.

In systems with a mobility edge~\cite{MAA,Mukerjeereview}, only the localized states contribute to
long time imbalance. The one particle Green's functions,
projected on these states, decay exponentially with distance. This defines an ``effective'' localization
length. The imbalance, divided by the fraction of localized states in
the system, is given by the same
analytic relations with this effective localization length. This
leads to non-analyticites in the imbalance as a function of
disorder strength, when the mobility edge passes through a band edge.

 Finally, we consider imbalance dynamics in a  Bose Hubbard
 model with an incommensurate potential. We use
a conserving approximation~\cite{conserving_KadanoffBaym}, keeping the lowest order processes leading
to dissipative and stochastic dynamics. Naively one would expect the memory of the initial
state to decay, as the Green's functions which propagate this memory
decay in time. However, as the quasiparticles decay, they create
excitations which act as a bath for the rest of the
quasiparticles. The noise fluctuations of this bath remembers the
initial conditions at strong disorder, and sustain the finite long
time imbalance. 

Imbalance dynamics in MBL systems has been treated theoretically using exact
diagonalization(ED), DMRG~\cite{MBLexpt1,DasSarmadyn} and in Hartree-Fock
approximation~\cite{HartreeFock}. However, ED and DMRG does not provide insight about the
mechanism that sustains the imbalance in interacting systems, while the
Hartree Fock approximation ignores the dissipative and stochastic
processes included in this work.

{\it Imbalance and Localization Length:} We consider non-interacting particles with 
\beq
H= -J\sum_{\langle ij\rangle}  a^\dagger_ia_j+\sum_i v(i) a^\dagger_ia_i,
\label{ham:ni}
\eeq
where $a^\dagger_i$ is the
particle creation operator on lattice site $i$, $J$ is the nearest
neighbour hopping and
$v(i)$ is a local potential. 
 \begin{figure*}
\includegraphics[width =0.95\linewidth]{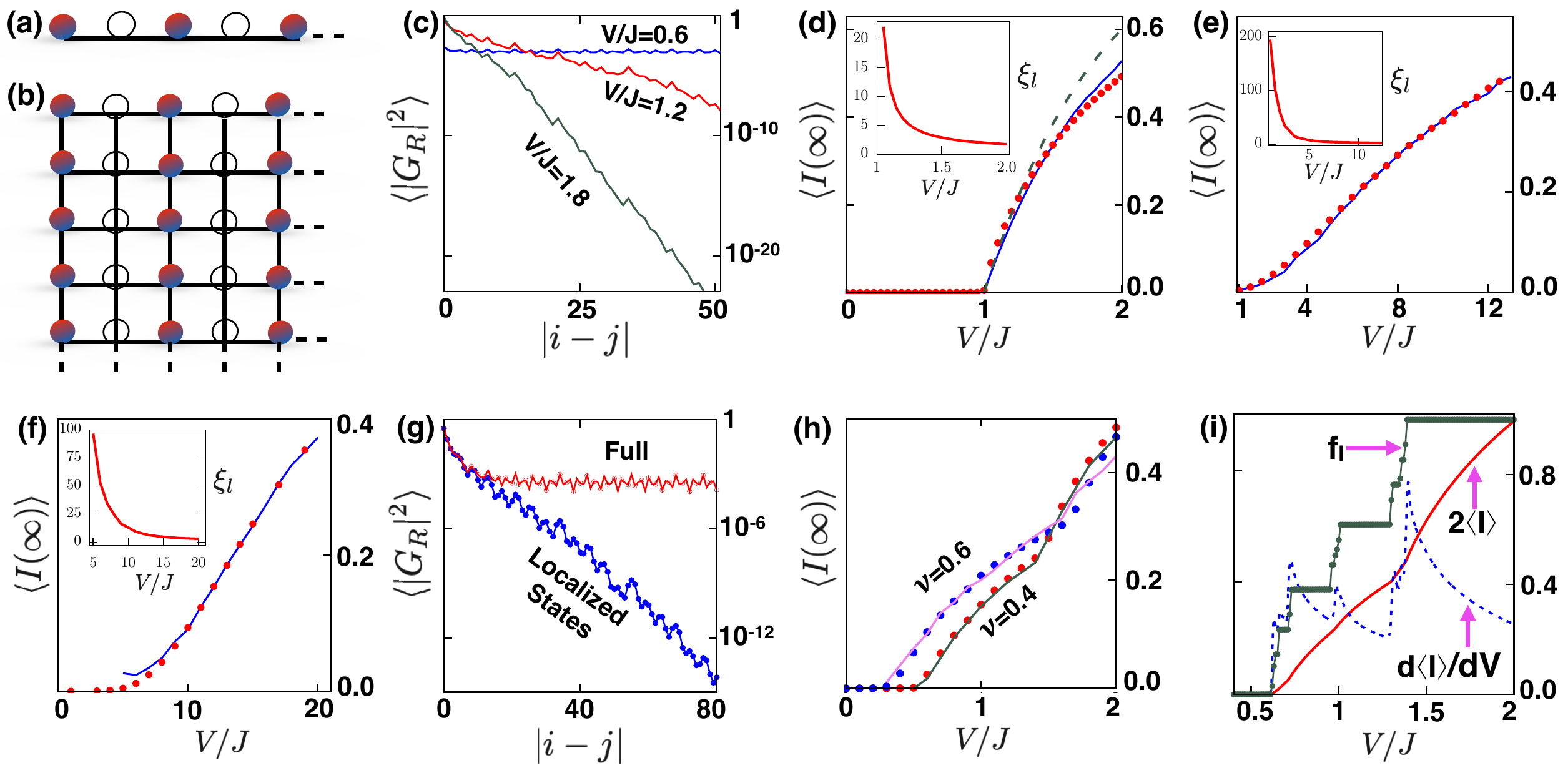}
 \caption{ Initial density profiles for imbalance dynamics in (a)
   linear chain and (b) square lattice. The solid(hollow) dots are
   particles (vacancies). (c) $\langle |G^\infty_R(i,j)|^2\rangle$ for
   Aubry Andre model as a
   function of $|i-j|$. For $V/J =0.6$, it saturates to a
   finite value; for $V/J=1.2,1.8$ it decays exponentially. (d)-(f):
   Long time imbalance $\langle I(\infty)\rangle$ as function of $V/J$ obtained
   from Eq.~\ref{imbalance:2} (solid dots) for (d) Aubry Andre model, (e) 1-d
   Anderson model and (f) square lattice Anderson model. The solid
   lines are Eq.~\ref{imb:tanh} (d,e)
   or Eq.~\ref{imb:square} (f) with $\xi_l$ obtained from fitting $|G_R|^2$ (see inset for $\xi_l$
   vs $V/J$). The dashed line in (d) uses $\xi_l/a=  log[V/J] $.
(g) $\langle |G^\infty_R|^2\rangle$ in the modified Aubry Andre model with mobility
   edge as a function of
 distance ($V/J=1.2$ and $\nu=0.4$). The full Greens function (which saturates) and its projection onto the
 localized states (which decay exponentially) are both shown. (h)
 $\langle I(\infty)\rangle$ for the modified Aubry Andre model as a
 function of $V/J$ for $\nu=0.4,0.6$ obtained from (i) Eq.~\ref{imbalance:2}
 (solid dots) (ii) Eq.~\ref{imb:mobedge}(solid lines). $\xi_l$ is
 obtained from fits of localized contributions in $|G_R|^2$. (i) The fraction of localized states
($f_l$), the long time imbalance $\langle I\rangle$(multiplied by $2$
to plot on same scale)and the derivative $d \langle I\rangle/dV$ as
a function of $V/J$ for the modified Aubry Andre model
($\nu=0.3$). The values of $V/J$, where the mobility edge leaves or
enters a band are marked by derivative discontinuities in $f_l$ and
$\langle I\rangle$. All data are averaged over $100$ disorder configurations. 1-d data are for $1000$ sites and
2-d data is for a $100 \times 100$ lattice. }
 \label{fig:imb_ni}
\end{figure*}

We will study dynamics of
this system within the Schwinger Keldysh field
theory~\cite{Kamenevbook}, which has two independent one particle correlators: (a) the
retarded Green's function, $G_R(i,t;j,t')$, which is the amplitude of propagating a particle to site $i$ at time $t$
provided the particle was at site $j$ at time $t'$, without creating
additional excitations, and (b) the Keldysh
Green's function $G_K(i,t;j,t')$, which represents the actual
amplitude of exchanging a particle between site $i$ at
time $t$ and site $j$ at time $t'$. $G_K(i,t;j,t)$ is related to densities and currents in the system; e.g. for
bosons (fermions), the local density $n_i(t)=(\pm 1/2)[i G_K(i,t;i,t)-1]$.

The system is initialized to a Fock state, where
$n_i(0)=1/2(1+\sigma_i)$, with $\sigma_i =\pm 1$ if $i\in A
(\bar{A})$. We use an extension of Keldysh field
theory, which can explicitly keep track of
arbitrary initial conditions in quantum dynamics ~\cite{chakraborty2}. Here, $G_R(i,t;j,t')=\sum_{n} \phi^\ast_n(i)\phi_n(j)
e^{-i E_n(t-t')}$, where $E_n$ and $\phi_n(i)$ are the energy levels
and corresponding wavefunctions. The Keldysh Green's function $G_K$ carries the
information of the initial density matrix and is given by
\bqa
\nonumber i G_K(i,t;j,t')&=&\sum_{k} G_R(i,t;k,0)G_R^\ast(j,t';k,0)
[1\pm 2n_k(0)],\\
n_i(t)&=& \sum_{k} |G_R(i,t;k,0)|^2 n_k(0).
\label{den:ni}
\eqa
Note that the expression for local density is same for bosons and
fermions. Hence, all the statements about imbalance dynamics in
non-interacting systems are independent of statistics of particles.
For a closed system
with equal number of  $A$ and $\bar{A}$ sites, the density imbalance, averaged over disorder, is 
\beq
\langle I(t)\rangle =\frac{2}{N}\sum_{ k\in
    A,i}\sigma_i\langle |G_R(i,t;k,0)|^2\rangle.
\label{imbalance:2}
\eeq
The Green's functions can be calculated from the knowledge
of energy eigenfunction in each disorder configuration, yielding a
``numerical'' estimate of imbalance.
$G_R(i,t;k,0) $ is the solution to Anderson's original problem~\cite{AL_original}: it is the
wavefunction at site $i$ and time $t$ of a particle initially
localized at $k$.  As $t\rightarrow \infty$, in the localized
phase, $\langle |G^\infty_R(i,k)|^2\rangle =\langle\sum_n |\phi_n(i)\phi^\ast_n(k)|^2\rangle \sim
e^{-2|r_i-r_k|/\xi_l}$. This decay defines the localization length
$\xi_l$. The long time imbalance 
\beq
\langle I(\infty)\rangle =\frac{2}{N}\sum_i\sum_{ k\in
    A}\sigma_ie^{-\frac{2|r_i-r_k|}{\xi_l}}. 
\label{imbalance:3}
\eeq
This is the universal relation between the long time imbalance, the localization length and the geometry of the initial
pattern. We will now consider some specific initial patterns that have been used in the cold
atom experiments on MBL~\cite{MBLexpt1,MBLexpt4}.

{\it 1-d chain with alternating pattern:} We consider a linear chain
with an initial state which has alternating $|1,0,1,0,... \rangle$
pattern ~\cite{MBLexpt1}, as shown in Fig.~\ref{fig:imb_ni}(a). Assuming a large chain
where boundary effects can be neglected (see SM for details), the sum
in Eq.~\ref{imbalance:3} gives
\beq
\langle I(\infty)\rangle = \tanh
\left(\frac{a}{\xi_l}\right).
\label{imb:tanh}
\eeq
From this ``analytic'' estimate, we see that as $\xi_l/a \rightarrow
\infty$, $\langle I(\infty)\rangle
\sim \frac{a}{\xi_l} \rightarrow 0$; 
i.e. (i) the memory retention is 
related to localization and (ii) close to a
localization-delocalization transition, the scaling of the Lyapunov exponent~\cite{Thouless1983} $\gamma = a/\xi_l$ governs the behaviour of the
density imbalance.

We first consider the Aubry Andre model~\cite{AA_original}, with an incommensurate $v(i) =V \cos[ 2\pi \alpha i
+\theta]$, where $\alpha=(\sqrt{5}+1)/2$ is the golden mean, and $\theta$
is a uniformly distributed random phase. This model, which is
implemented in cold atom experiments~\cite{MBLexpt1}, has a
localization-delocalization transition at $V/J=1$~\cite{AA_original}. This can be clearly
seen in Fig.~\ref{fig:imb_ni}(c), where we plot $\langle
|G^\infty_R(i,j)|^2\rangle$ as a function of $|i-j|$. For $V/J=0.6$,
where the system is delocalized, $\langle |G^\infty_R|^2\rangle$ saturates to a finite
value at large distances, whereas it shows an exponential decay for
$V/J >1$.  In Fig.~\ref{fig:imb_ni} (d), we plot the long time
imbalance as a function of $V/J$ obtained using
the numerical estimate from Eq.~\ref{imbalance:2} (solid dots) . We
also plot the analytic answer from  Eq.~\ref{imb:tanh} with
$\xi_l$ obtained from (i)  fitting $\langle |G^\infty_R|^2\rangle$ (solid
line)[see inset for $\xi_l$ vs $V/J$] and (ii) a duality relation ~\cite{AA_duality} $\xi_l =
a~ log[V/J]$ (dashed line). The analytic answer matches the
numerical estimate for $\xi_l/a > 1$.
We also consider the 1-d Anderson model~\cite{AL_original}
where each $v(i)$ is an independent random variable, with $P[v(i)]=\Theta[V^2/4-v^2(i)]1/V$. This system is localized for any $V/J$,
with a localization length $\xi_l/a\sim (V/J)^{-2}$~\cite{illcondmat} for weak disorder
(see inset of Fig.~\ref{fig:imb_ni} (e)). In Fig.~\ref{fig:imb_ni} (e)
we plot the long time imbalance obtained from Eq.~\ref{imbalance:2} (solid dots) and Eq.~\ref{imb:tanh}(solid
line) and find good quantitative match between these estimates.
\begin{figure}
 \includegraphics[width=0.95\linewidth]{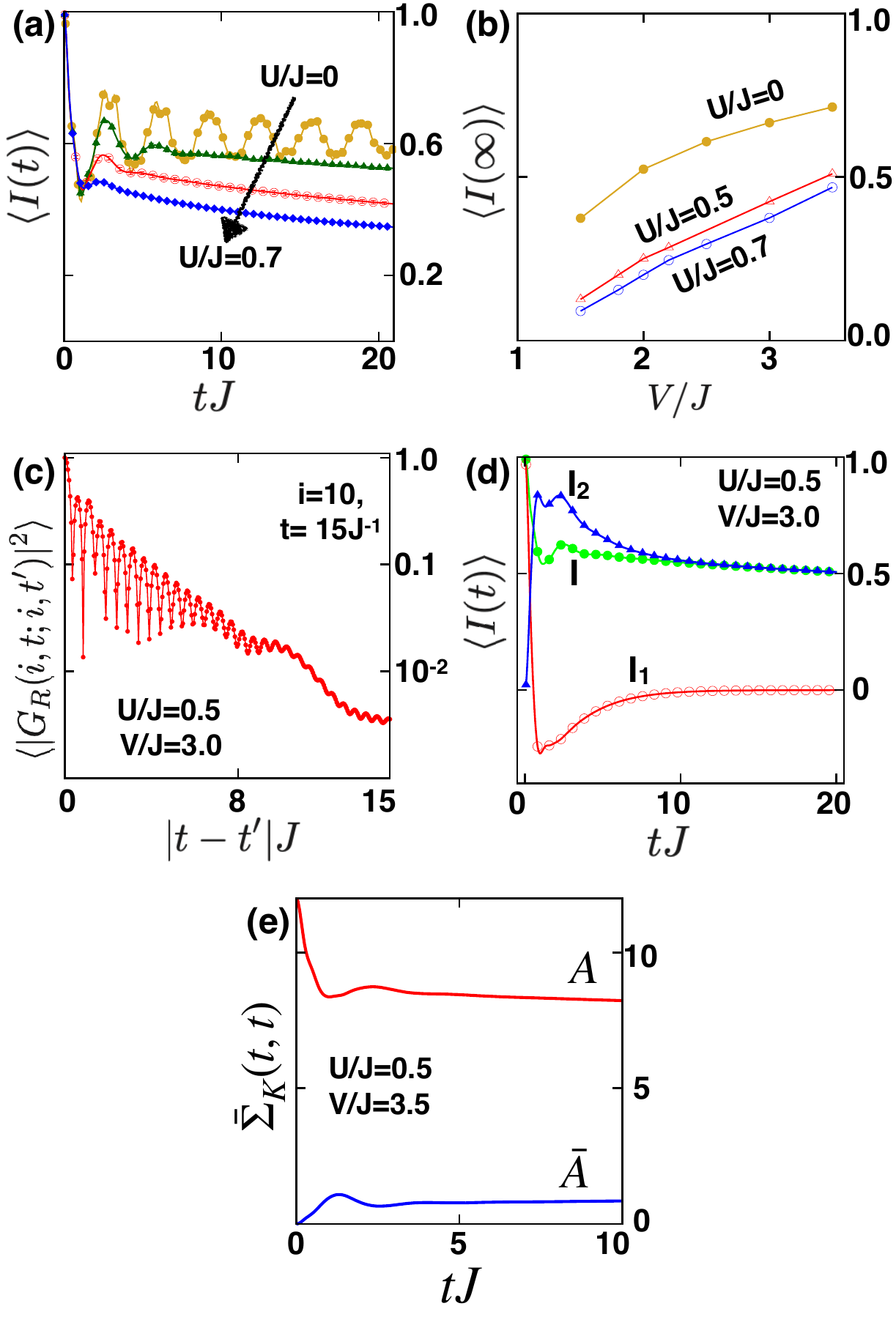}
 \caption{(a) $I(t)$ for Bosons in an
   Aubry Andre Hubbard model with $V/J=2.5$ and
   $U/J=0.0,~0.3,~0.5,~0.7$. For finite $U$, the imbalance decays
   exponentially to a finite value. (b) The long time
   imbalance $\langle I(\infty)\rangle$ as function of $V/J$ for
   different values of $U/J$. (c) The exponential decay in time for
   $G_R$ due to dissipative process. (d) The contribution to imbalance
 due to (i) the direct decay of initial correlations $I_1$ and (ii)
 stochastic fluctuation due to effective bath $I_2$. The long time
 imbalance is dominated by $I_2$.(e) The space-time local part of
 $\Sigma_K$, averaged over $A$ and
 ${\bar A}$ sites. The bath clearly
 distinguishes between $A$ and ${\bar A}$ sites at long times.  All data in (a)-(d) are for $N=20$ sites and
 averaged over $50$ configurations. (e) is obtained from $N=10$ site system averaged over $200$ samples.}
\label{fig:imb_int}
 \end{figure}

{\it Disordered Square Lattice:} We consider the Anderson model of
uniformly distributed random potentials on a
square lattice~\cite{AL_original}. This model is localized for all $V/J$, with $\langle |G^\infty_R(r)|^2\rangle \sim
e^{-2\sqrt{r_x^2+r_y^2}/\xi_l}$, and $\xi_l \sim a ~e^{J^2 /V^2}$~\cite{illcondmat}. We
consider the experimentally relevant \cite{MBLexpt2,MBLexpt4} initial density pattern of alternating chains
which have $1$ and $0$ particles on each site, as shown in
Fig~\ref{fig:imb_ni}(b). The long time imbalance 
\beq
\langle I(\infty)\rangle =\sum_{n_x,n_y=-\infty}^\infty (-1)^{n_x}
e^{-\frac{2a}{\xi_l} \sqrt{n_x^2+n_y^2}}.
\label{imb:square}
\eeq  
While this sum cannot be done analytically, numerical evaluation (see
SM for details) shows $\langle I(\xi_l)\rangle \sim (\sqrt{32}/\pi^2)(a/\xi_l)$ for
$\xi_l \gg a$. In Fig~\ref{fig:imb_ni}(f), we plot the imbalance obtained
from Eq.~\ref{imbalance:2} (solid dots) together with that obtained
from Eq.~\ref{imb:square} (solid lines), with $\xi_l$ (see inset) obtained from
exponential fit of the $\langle |G^\infty_R|^2\rangle$. The two approaches
match till $V/J=5$, when $\xi_l \sim 100 a$ and our $100\times 100$ system is effectively
delocalized, The cold atom
experiments, which are restricted to similar sizes, may also
see effective delocalization at this scale.

{\it Mobility Edges and Imbalance:} We now turn our attention to the
modified Aubry Andre model~\cite{MAA,Mukerjeereview} in 1d, where $v(i) =V \cos[ 2\pi \alpha i
+\theta]/ (1-\nu \cos[ 2\pi \alpha i
+\theta])$, with $-1< \nu < 1$. $\nu=0$ corresponds to the Aubry
Andre model discussed before.  At low $V/J$ the model has delocalized states, for
intermediate values of $V/J$ it supports a mobility edge at $E_c=2(J-V)/\nu$~\cite{MAA}, which is an
energy threshold separating localized and delocalized states. At large
$V/J$ all states are localized in the system. 
In
Fig.~\ref{fig:imb_ni} (g), we plot $\langle |G^\infty_R|^2\rangle$ of the
system as a function of distance for $V/J=1.2$ and $\nu=0.4$, where there is a
mobility edge. The long distance behaviour of $\langle |G^\infty_R|^2\rangle$ is
dominated by delocalized states and saturates to a constant. In the
same figure, we also plot the contribution to $\langle |G^\infty_R|^2\rangle$ from states
above the mobility edge, which clearly shows an exponential
decay. This decay can be used to extract an ``effective'' localization
length $\tilde{\xi}_l$ for
the system. The imbalance in this case is given by
\beq
\langle I(\infty)\rangle = f_l \tanh
\left(\frac{a}{\tilde{\xi}_l}\right),
\label{imb:mobedge}
\eeq
where $f_l$ is the fraction of localized states. In
Fig.~\ref{fig:imb_ni} (h), we plot $\langle I(\infty)\rangle$ as a
function of $V/J$ for $\nu=0.4,0.6$. The solid dots
(Eq.~\ref{imbalance:2}) and the solid lines (Eq.~\ref{imb:mobedge})
track each other. The imbalance goes to $0$
when all states are delocalized at low $V/J$. At large $V/J$, the
curve approaches the $\nu=0$ answer. 

There is a  clear
non-analytic feature of the imbalance as a function of $V/J$, which
coincides with the $V/J$ where the mobility edge coincides with the
band edge. A closer scrutiny shows that the system
has multiple bands and there is a sharp change in derivative every time
the mobility edge coincides with a band edge.  This can be seen
in  Fig.~\ref{fig:imb_ni} (i), where we plot $f_l$, $\langle I(\infty)\rangle$ and $d\langle I(\infty)\rangle /dV$ vs $V/J$ in the same
plot for $\nu=0.3$. To understand this non-analyticity, consider
the rightmost feature, at $V_0\sim 1.4 J$. For $V>V_0$, the
mobility edge is below the lowest band; all states are
localized and contribute to $I$. As we approach $V_0$, the singular
part of the imbalance is governed by the scaling of the energy
dependent localization
length, $\xi_l(E) \sim (E-E_c)^{-\beta}$, leading to $I_s\sim
(V-V_0)^\beta$. On the other hand, for
$V<V_0$, there is an additional effect as the fraction of localized
states also decrease. If the Van Hove singularity in the density of states at the band edge $E_b$,
$\rho(E) \sim (E-E_b)^{-\delta}$, the fraction of localized states
changes as $\Delta f_l\sim |V-V_0|^{1-\delta}$, and hence $I_s\sim |V-V_0|^{\beta+1-\delta}\sim
|V-V_0|^{2\beta} $.  Here we have used the well known formula $\beta=1-\delta$~\cite{DasSarma_Xie}. This
leads to the cusp like behaviour of $d\langle I(\infty)\rangle /dV$ in Fig.~\ref{fig:imb_ni} (i) when the mobility edge and band
edge coincide.

{\it Imbalance Dynamics in Interacting Systems:} Finally we focus our
attention on the Bose Hubbard model with Aubry Andre
potential in 1-d, where we add to the Hamiltonian of Eq.~\ref{ham:ni}
the local Hubbard repulsion $U\sum_i n_i(n_i-1)$.
The interaction effects on the Greens functions are incorporated through
retarded and Keldysh self-energies, $\Sigma_R$ and $\Sigma_K$, where the imaginary part of
$\Sigma_R$ is related to the dissipation in the system and $\Sigma_K$
is the noise correlator due to the effective bath formed by the
medium~\cite{chakraborty1}. The interacting Green's functions are obtained from~\cite{Kamenevbook,chakraborty2} 
\begin{widetext}
\bqa
\label{grfn:int}
G_R(i,t;j,t')&=&G_{R0}(i,t;j,t')+\int_{t'}^{t}dt_1\int_{t'}^{t_1}dt_2
G_{R0}(i,t;k,t_1)\Sigma_R(k,t_1;l,t_2) G_R(l,t_2,j,t')\\
\nonumber G_K(i,t;j,t')&=&-i~
G_R(i,t;k,0)[1+2n_k(0)]G_R^\ast(j,t';k,0)+\int_{0}^{t}dt_1\int_{0}^{t'}dt_2
G_{R}(i,t;k,t_1)\Sigma_K(k,t_1;l,t_2) G_R^\ast(j,t';l,t_2).
\label{selfcon_int}
\eqa
\end{widetext}
Here $G_{R0}$ is the non-interacting retarded
Green's function, and we have neglected connected many particle
correlations in the initial state. 
We work with a conserving approximation~\cite{conserving_KadanoffBaym}, where we
keep all skeleton diagrams upto second order in $U$ to calculate the
self-energies (see SM for details). Our approximation keeps the minimal
non-trivial diagrams which lead to
dissipative and stochastic dynamics in the system. The resulting
imbalance, plotted in Fig.~\ref{fig:imb_int}(a) for $V=2.5J$ and different
values of $U/J$, shows an exponential decay in
time, which can be fitted to $\langle I(t)\rangle =\langle
I(\infty)\rangle + \kappa e^{-\mu t}$. The long time imbalance $\langle
I(\infty)\rangle$, obtained from this fit, is plotted for different
$V/J$ in Fig~\ref{fig:imb_int}(b). The system can sustain a finite
  imbalance, although interaction reduces its value. We note that our
  calculation is likely to overestimate the effects of interaction,
  since we do not take into account screening of the bare interaction strength.

In the interacting system, as a particle propagates, it creates
additional excitations in the system by scattering. Since $G_R$ is the
amplitude of propagation {\it without creating additional
  excitations}, $\langle |G_R(t,0)|^2\rangle$ decays exponentially
with time, as shown in
Fig~\ref{fig:imb_int}(c). In Eq.~\ref{grfn:int} for $G_K$, the first
term is a modification of the non-interacting answer, with the initial
profile propagated by the interacting $G_R$. It is obvious that this
term decays to $0$ at long
times. However, the excitations created in the medium act as a bath
for the particle, and the stochastic fluctuations of this bath is represented by the second term. The
contributions of these terms to the density imbalance, $I_1$ and $I_2$
are plotted with time in Fig~\ref{fig:imb_int}(d). As expected, $I_1$
decays to zero at long times and $I_2$ dominates the finite
imbalance. The memory of the initial conditions now resides in the
noise correlators of the bath, which distinguishes between $A$ and
${\bar A}$ sites (see SM for details), and sustains the finite
imbalance. To see this, in Fig~\ref{fig:imb_int}(e), we plot the
space-time local part of the disorder averaged Keldysh self-energy
$\Sigma_K(i,t;i,t)$ (which is the effective variance of the local
noise fluctuations),
averaged over $A$ and ${\bar A}$ sites. $\bar{\Sigma}_K$ clearly
distinguishes between $A$ and ${\bar A}$ sites in the long time limit, which is key to
a finite imbalance in the system at long times.

We have used a recent extension of Keldysh field theory to provide
insight into how memory of initial states are retained in dynamics of
disordered systems. Considering experimental protocols in ultracold
atoms, we have derived exact relation between long time imbalance and localization length in
non-interacting systems. In interacting systems, our calculations show
that long time imbalance is sustained at strong disorder by the noise
correlations which remember the initial density pattern. 

\begin{acknowledgments}
The authors thank Eugene Demler and Sankar Das Sarma for useful discussions. The authors also acknowledge computational facilities at Department of Theoretical Physics, TIFR Mumbai. 
\end{acknowledgments}

\bibliographystyle{unsrt} \bibliography{Imbalance_2.bib}

\end{document}


\newcommand {\beq} {\begin{equation}}
\newcommand {\eeq} {\end{equation}}
\newcommand {\bqa} {\begin{eqnarray}}
\newcommand {\eqa} {\end{eqnarray}}
\newcommand {\ba} {\ensuremath{b^\dagger}}
\newcommand {\Ma} {\ensuremath{M^\dagger}}
\newcommand {\psia} {\ensuremath{\psi^\dagger}}
\newcommand {\psita} {\ensuremath{\tilde{\psi}^\dagger}}
\newcommand{\lp} {\ensuremath{{\lambda '}}}
\newcommand{\A} {\ensuremath{{\bf A}}}
\newcommand{\Q} {\ensuremath{{\bf Q}}}
\newcommand{\kk} {\ensuremath{{\bf k}}}
\newcommand{\qq} {\ensuremath{{\bf q}}}
\newcommand{\kp} {\ensuremath{{\bf k'}}}
\newcommand{\rr} {\ensuremath{{\bf r}}}
\newcommand{\rp} {\ensuremath{{\bf r'}}}
\newcommand {\ep} {\ensuremath{\epsilon}}
\newcommand{\nbr} {\ensuremath{\langle ij \rangle}}
\newcommand {\no} {\nonumber}
\newcommand{\up} {\ensuremath{\uparrow}}
\newcommand{\dn} {\ensuremath{\downarrow}}
\newcommand{\rcol} {\textcolor{red}}

\title{Supplementary Material for "Memories of initial states and density imbalance in dynamics of disordered systems"}
\author{Ahana Chakraborty}\email{ahana@theory.tifr.res.in}
 \affiliation{Department of Theoretical Physics, Tata Institute of Fundamental
 Research, Mumbai 400005, India.}
\author{ Pranay Gorantla} 
\affiliation{Department of Theoretical Physics, Tata Institute of Fundamental
 Research, Mumbai 400005, India.}
\affiliation{Department of Physics, Princeton University, Washington Road,
Princeton, NJ 08544, USA}
\author{Rajdeep Sensarma}
 \affiliation{Department of Theoretical Physics, Tata Institute of Fundamental
 Research, Mumbai 400005, India.}

\pacs{}
\date{\today}

\maketitle

\appendix

\section{Finite size correction in long time imbalance for $1-d$ chain with alternating pattern}
\begin{figure}[h]
 \includegraphics[width=0.5\linewidth]{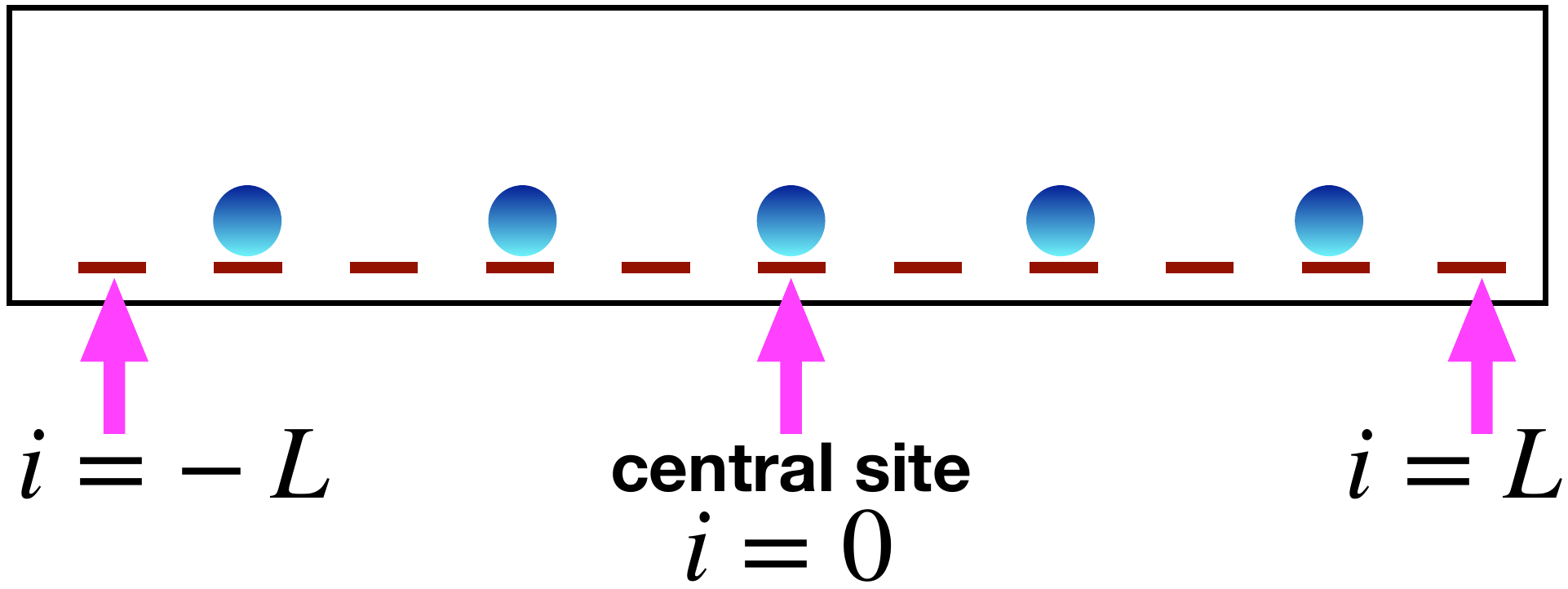}
 \caption{Linear chain consisting of odd number of sites, $2L+1$ with alternating density pattern. The central site is chosen to belong to filled sub-lattice $A$. }
\label{fig:finitesize}
 \end{figure}
In this appendix, we will work out the analytical formula for the long
time imbalance, $\langle I(\infty) \rangle$ in case of a finite size linear chain initialized in the state with alternating density pattern shown in Fig.\ref{fig:finitesize}. We will show that the finite size correction to the analytical formula for the longtime imbalance, $\langle I(\infty) \rangle=\tanh(a/\xi_l)$ is exponentially small in the system size. To show this, we consider a linear chain having $2L+1$ number of sites. We choose $L$ to be an odd number and the origin, situated at the central site of the chain,belong to the filled sub-lattice $A$. We note that the final conclusion about exponentially small finite size correction to $\langle I(\infty) \rangle$ does not depend on this particular choice of the geometry. Using equation 4 of the main text, we write the longtime imbalance of the finite chain as,
\bqa
\langle I(\infty) \rangle &=& \frac{1}{L} \left[\sum \limits_{i=0,\pm 2,...,\pm(L-1)} ~~\sum \limits_{j=0,\pm 2,...,\pm(L-1)} e^{-2\frac{|i-j| ~a}{\xi_l}} -
\sum \limits_{i=0,\pm 2,...,\pm(L-1)} ~~\sum \limits_{j=\pm 1,\pm 3,...,\pm L} e^{-2\frac{|i-j|~a}{\xi_l}}
\right]  \nonumber \\
&=&\left[2 \left ( \sum \limits_{r=0, 2,...,(L-1)} e^{-2\frac{r~a}{\xi_l}} -
\sum \limits_{r= 1, 3,...,L} e^{-2\frac{r~a}{\xi_l}} \right ) -1
\right] \nonumber \\
&=& \tanh \left(\frac{a}{\xi_l} \right)-\frac{2}{1+e^{-\frac{2~a}{\xi_l}}} e^{-\frac{2(L+1)~a}{\xi_l}},
\eqa
where $a$ is the lattice spacing. In this equation, the second term gives the finite size correction to the longtime imbalance, $\langle I(\infty) \rangle$. In the large system size limit, $L~a/\xi_l >>1$, the correction term exponentially decays to zero and we recover Eq.5 of the main text.

\section{Retarded Green's functions and localization length for
  Anderson model on a linear chain}
\begin{figure}[t]
 \includegraphics[width=0.65\linewidth]{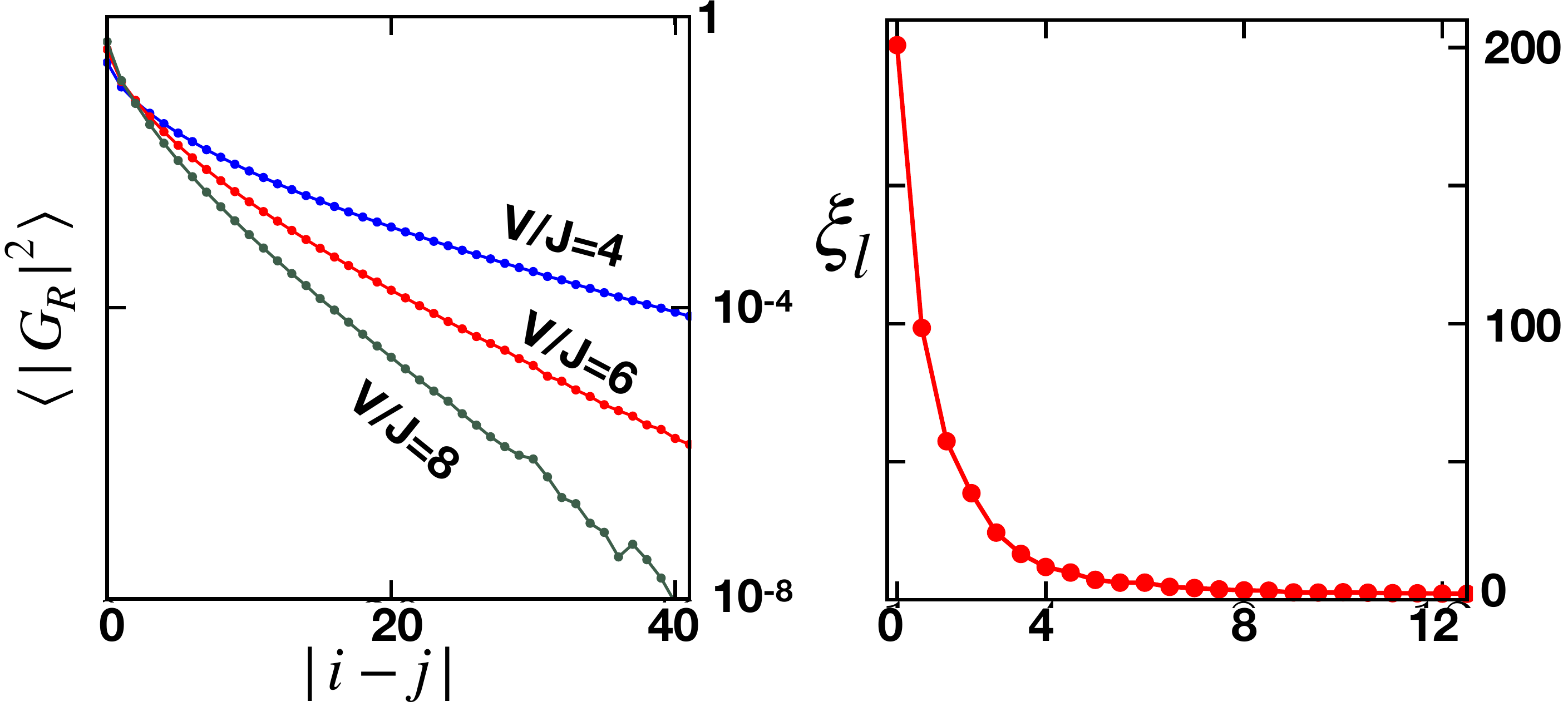}
 \caption{(a) Exponential decay of long time disorder averaged Green's
   function, $\langle |G^\infty_R(i,j)|^2\rangle$ with $|i-j|$ for the
 1-d Anderson model for $V/J=4,6,8$ (b) The localization length,
 extracted from the exponential fit , as a function of $V/J$. Data is
 for $N=1000$ lattice, averaged over $100$ disorder realizations.}
\label{fig:AL}
 \end{figure}
 In the main text, we have shown that the longtime imbalance retained
 by the system is related to the retarded Green's function as, 
$\langle I(t)\rangle =\frac{2}{N}\sum_{ k\in
    A,i}\sigma_i\langle |G_R(i,t;k,0)|^2\rangle$. Here
  $|G_R(i,t;j,t')|^2$ represents the probability of finding a particle
  at a site $i$ at time t, given that the particle was at site $j$ at
  some previous time $t'$. It has been shown in
  Ref. \onlinecite{chakraborty2} that $G_R(i,t;j,t')$ does not depend
  on the initial condition of the system and hence it is insensitive
  to the initial density imbalance imprinted on the system between $A$
  and $\bar{A}$ sub-lattices. The system recovers translation
  invariance when one looks at disorder averaged correlation
  functions, hence the disorder averaged $\langle |G_R(i,t;k,t')|^2\rangle$ becomes function of only $|r_i-r_j|$. In the Anderson \cite{AL_original} model of random disorder potential on a linear chain,  it is well established that all the eigenstates becomes localized in space in presence of infinitesimal disorder strength, $V$. In this case, $\langle |G_R(i,t;k,0)|^2\rangle$ in the longtime limit also shows an exponential decay, $\langle |G^\infty_R(i,k)|^2\rangle =\langle\sum_n |\phi_n(i)\phi^\ast_n(k)|^2\rangle \sim
e^{-2|r_i-r_k|/\xi_l}$ with localization length $\xi_l$. In Fig.\ref{fig:AL}, we have plotted $\langle |G^\infty_R(i,k)|^2\rangle$ as a function of $|i-k|$ in a semi-log plot for three different disorder strength $V/J=4,6,8$. Localization length, $\xi_l$, shown in Fig.\ref{fig:AL} as a function of $V$, is extracted from the slope of the exponential decay of $\langle |G^\infty_R|^2\rangle$ with distance. These $\xi_l$ for different $V$ are used to calculate the analytical longtime imbalance, $\langle I(\infty) \rangle=\tanh(a/\xi_l)$, shown by solid line in the Fig.1 of the main text.

\section{Green's functions and localization length for square lattice
  Anderson model}
\begin{figure}[h]
 \includegraphics[width=0.85\linewidth]{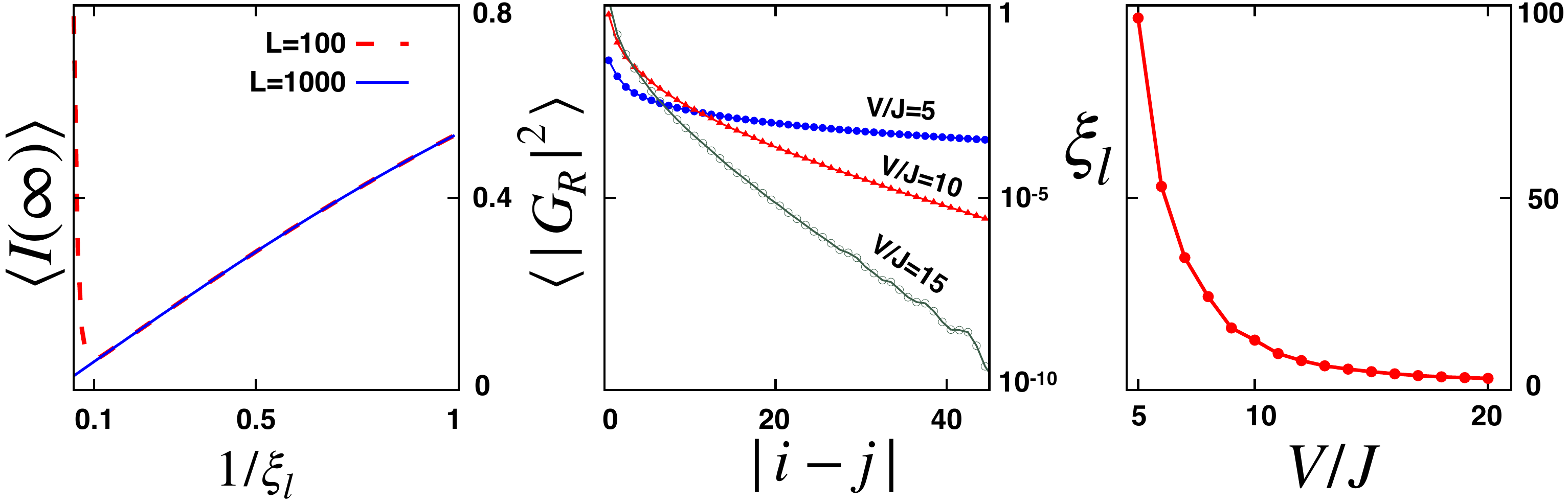}
 \caption{(a) Numerical evaluation of the analytic formula for long
   time imbalance in the 2d  localized phase, as a function of the inverse localization
   length. The two plots are evaluating the formula in a lattice of
   size $N=100 \times 100$ and $N=1000\times 1000$. Note the linear behaviour in both
   curves. The upturn in the $N=100$ curve is due to finite size effects. (b) Exponential decay of long time disorder averaged Green's
   function, $\langle |G^\infty_R(i,j)|^2\rangle$ with $|i-j|$ for the
 2-d Anderson model for $V/J=5,10,15$ (b) The localization length,
 extracted from the exponential fit , as a function of $V/J$. Data is
 for $100\times 100$ lattice averaged over $100$ realizations. Note
 that at $V/J=5$, the extracted localization length $\xi_l \sim 100$
 and the system is effectively delocalized. }
\label{fig:AL2D}
 \end{figure}
In the main text we had obtained the long time imbalance in a square
lattice for a system starting with initial configuration shown in
Fig~1(b). In terms of the localization length we
obtained 
\beq
\langle I(\infty)\rangle = \sum_{n_x=-\infty}^\infty
\sum_{n_y=-\infty}^\infty (-1)^{n_x} e^{-\frac{2a}{\xi_l}\sqrt{n_x^2+n_y^2}}
\eeq
In Fig.~\ref{fig:AL2D}(a), we plot the numerical evaluation of this sum
as a function of $a/\xi_l$. The limits of the sum has been set to
$[-L,L]$, and we show plots for $L=100$ and $L=1000$. The graphs
follow a straight line, showing that the imbalance scales with the
inverse localization length. Numerically we find the slope to be $\sqrt{32}/\pi^2$. For $L=100$, we see a sharp upturn at
$a/\xi_l \sim 0.1$, where finite size effects start to play a
role. For $L=1000$, the curve continues with the same slope. Thus (a)
finite size effects are easy to detect in this sum and (ii) the 
slope calculations from such finite size sums are reliable as long as
we stay away from the upturn in the curve.

The Anderson model on a square lattice is localized for arbitrary
small disorder strength. However, since the localization length
grows logarithmically at weak disorder, finite size systems will be
effectively delocalized over a range of $V/J$. In Fig.~\ref{fig:AL2D}(b),
we plot the disorder averaged retarded Green's function for the
Anderson model in 2d in the long time limit, using the formula,
$|G^\infty_R(i,j)|^2=\sum_n |\phi_n(i)\phi_n(j)|^2$, where the sum is
over eigenstates of the single particle Hamiltonian and $\phi_n(i)$ is
the corresponding wavefunction. We plot it for $V=4,6,8$, all of which
show exponential decay. The localization length calculated from this
decay is plotted in Fig.~\ref{fig:AL2D}(c) as a function of $V/J$. We
note that for $V/J=5$, $\xi_l\sim 100 a$. Since we are working with a
$100 \times 100$ lattice, the estimates here are unreliable. Reliable
estimates can be obtained in this case for $V/J >10$, which corresponds
to a $\xi_l=10 a$, where sharp upturns are seen in
Fig.~\ref{fig:AL2D}(a).

\section{Interacting Disordered systems}
\begin{figure}[h]
 \includegraphics[width=0.7\linewidth]{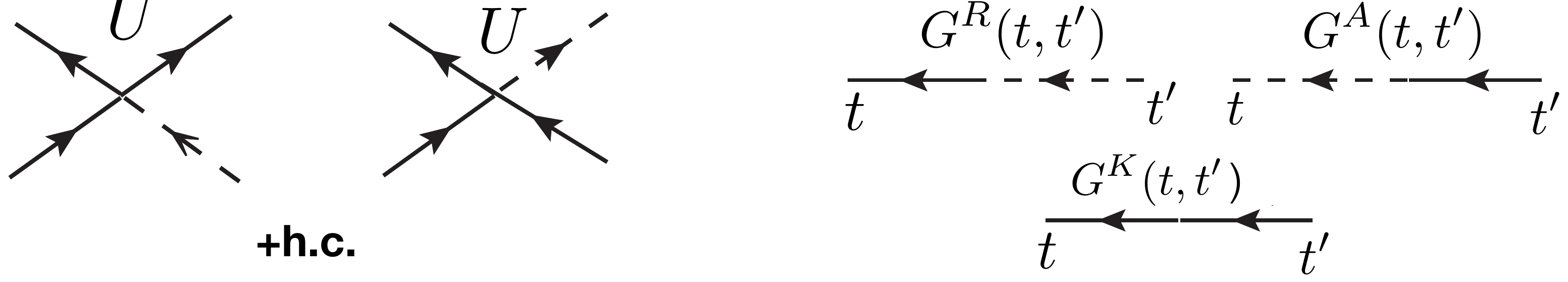}
 \caption{The Feynman interaction vertices and the symbolic
   propagators for a system of interacting bosons with a local Hubbard
 repulsion $U$.}
\label{fig:vertex}
 \end{figure}
In this section we provide the details of the imbalance dynamics for
interacting bosons in 1d. We work with the Hamiltonian
\beq
\sum_i -Ja^\dagger_ia_{i+1} +v(i)a^\dagger_ia_i +Un_i(n_i-1)
\eeq
where $a^\dagger_i$ is the boson creation operator on site $i$,$n_i$
is the number operator on site $i$. Here $J$ is the hopping, $U>0$ is
the Hubbard repulsion and $v(i)=V\cos(2\pi\alpha i +\theta)$ is the
Aubry Andre potential with $\alpha =(\sqrt{5}-1)/2$ and $\theta$ a
random phase with uniform distribution. 
\begin{figure}[h]
 \includegraphics[width=0.7\linewidth]{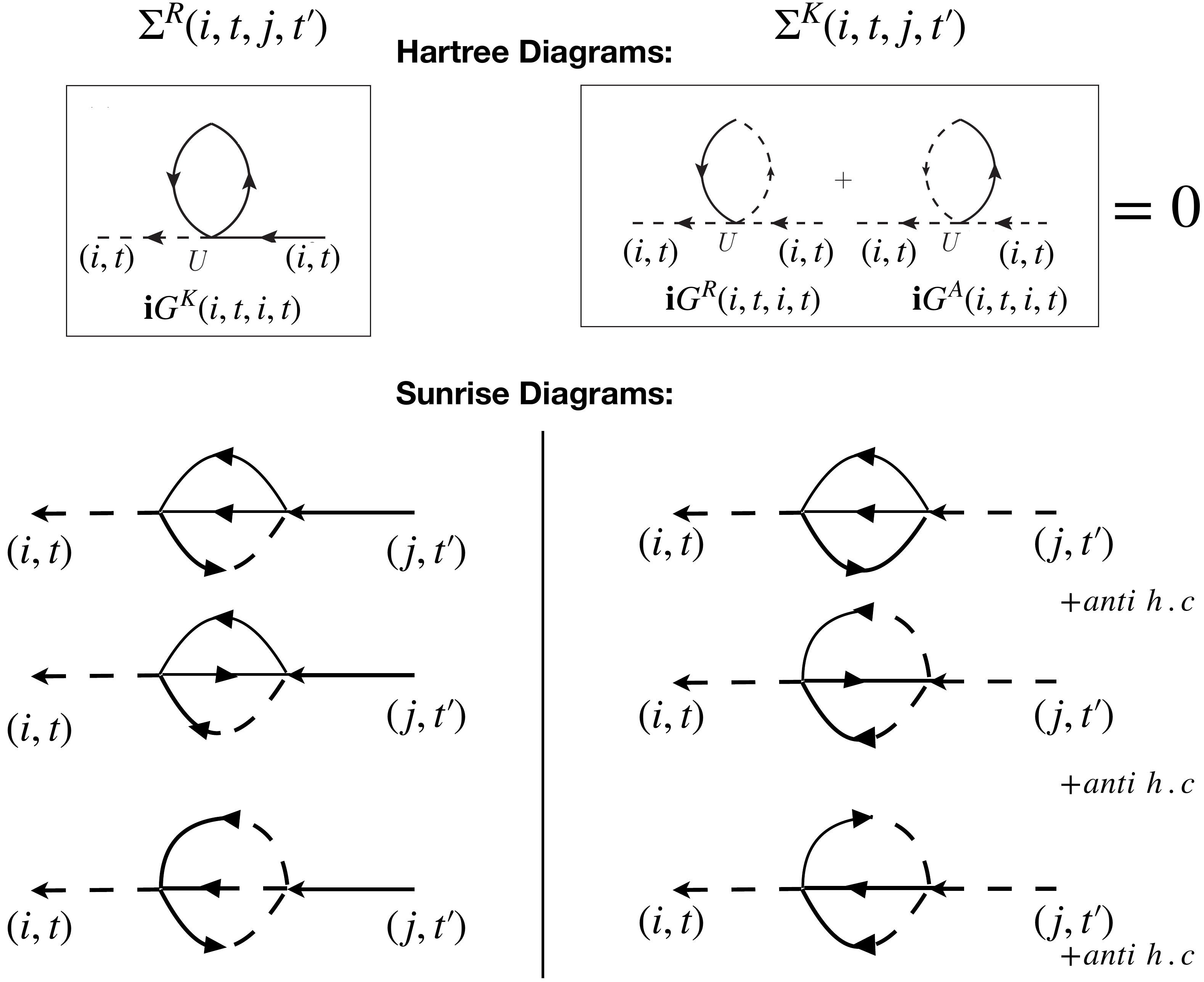}
 \caption{The Feynman diagrams used to evaluate self energies in the
   Bose Hubbard model with the Aubry Andre potential. The Green's
   functions are all interacting Green's functions, so that a
   conserving approximation is obtained. The self energies contain
   skeleton diagrams upto second order in $U$. }
\label{fig:FeynSig}
 \end{figure}
The effect of interactions is incorporated through the retarded and
Keldysh self energies, $\Sigma_R$ and $\Sigma_K$. The physical meaning
of the self energies become apparent from the classical
saddle point equation of motion for the boson fields
\beq
[i\partial_t -v(i)]\phi_i(t) +J\phi_{i\pm 1}(t) -\int dt'
\Sigma_R(i,t;j,t') \phi_j(t')=\eta_i(t)
\eeq
where the random noise $\eta$ has correlators $\langle
\eta_i(t)\eta_j(t')\rangle =i\Sigma_K(i,t;j,t')$. The real and
imaginary part of $\Sigma_R$ correspond to the modification of the
Hamiltonian and dissipation in the system.

We note that we have ignored the effects of initial connected
correlations in writing a self energy. We work with number conserving
approximations by constructing skeleton expansions for the self
energies. The self energies are then functions of the interacting
Green's functions, i.e. one has to solve the integral equations
Eq.~9 of the main text  selfconsistently, with $\Sigma=\Sigma(G)$.

We keep all skeleton diagrams upto second order in $U$ in our
calculation. See Fig.~\ref{fig:FeynSig} for the details of the
diagrams. The self energies are then given by
\bqa
\nonumber \Sigma_R(i,t;j,t')&=& 2\delta_{ij}\delta(t-t') (i
U)G_K(i,t;i,t)-2U^2[2G_K(i,t,j,t') G_K(j,t',i,t) G_R(i,t,j,t')+\nonumber \\ &&G_K(i,t,j,t') G_A(j,t',i,t) G_K(i,t,j,t')+G_R(i,t,j,t') G_A(j,t',i,t) G_R(i,t,j,t')] \nonumber \\
\Sigma_K(i,t;j,t')&=& -2 U^2[ 2G_K(i,t,j,t') G_A(j,t',i,t) G_R(i,t,j,t') + \nonumber \\
&~&G_K(j,t',i,t) G_R(i,t,j,t') G_R(i,t,j,t') + G_K(j,t',i,t) G_K(i,t,j,t') G_K(i,t,j,t') ]
\eqa
We note that the approximation is non-perturbative in $U$,
since the Green's functions used in the diagrams are the full
interacting $G$s. The first order diagram leads to the self-consistent
Hartree approximation. At this order, $\Sigma_K=0$ and $\Sigma_R$ is
real. Thus there is no dissipation or noise in the dynamics of the
system, the dynamics is controlled by time dependent dephasing. The
second order sunrise diagrams lead to both dissipation and noise, so
that there is a possibility of thermalization in the system. We
numerically solve this large number of integral equations, maintaining
an error of $< 0.5\%$ in the number conservation.
%

\bibliographystyle{apsrev4-1}
\bibliography{Imbalance_2.bib}